\documentclass[12pt,a4 paper]{article}
\usepackage{amssymb}
\usepackage{amsmath}
\usepackage[dvips]{graphicx}
\usepackage{graphicx}
\begin{document}
\begin {center}
\textbf{ A note on characterization based on past entropy} \\
\vspace{.4 cm}
Richa Thapliyal and H.C.Taneja\\
Department of Applied Mathematics,\\
Delhi Technological University,\\
Bawana Road, Delhi-110042, India.\\
\end {center}
\begin{abstract}
Ebrahimi (1996) has shown that the measure of residual entropy characterizes the distribution function uniquely. In this communication we study an analogous result for past entropy.
\end{abstract}


 Let $X$ denote the lifetime of a system, a component or living organism with probability density function $f$, distribution function $F$ and survival function $\bar{F}=1-F$. Shannon (1948) defined a measure of uncertainty associated with $X$  given by
  \begin{equation}
  H(X)=-\int_{0}^{\infty}f(x)\log{f(x)}dx.
  \end{equation}
 The role of entropy in residual lifetime distributions has attracted attention in recent years. Ebrahimi (1996) defined uncertainty of the residual lifetime of the random variable $X_{t}=[X-t|X>{t}]$ as
  \begin{equation}
  H(f;t)=-\int_{t}^{\infty}\frac{f(x)}{\bar{F}(t)}\log{\frac{f(x)}{\bar{F}(t)}}dx
  \end{equation}
  where $t\geq{0}$ and $X>t$. For $t={0}$, (2) reduces to (1). Ebrahimi has also studied that the measure (2) characterizes the distribution function uniquely. Crescenzo and Longobardi (2002) introduced a measure to calculate uncertainty for past time distribution called past entropy, defined as
 \begin{equation}
 \bar{H}(X;t)=-\int_{0}^{t}\frac{f(x)}{F(t)}\log{\frac{f(x)}{F(t)}}dx=1-\frac{1}{F(t)}\int_{0}^{t}f(x)\log{\phi(t)}dx,
 \end{equation}
 where $\phi(t)$ or ${\phi}_{X}(t)$, for $t>{0}$, is the reversed hazard rate function given 
 \footnotetext[1]{$richa31aug@gmail.com$}
\footnotetext[2]{$hctaneja@rediffmail.com$}
 by
 \begin{equation}
 {\phi}_{X}(t)=\phi(t)=\frac{d}{dt}\log{F(t)}=\frac{f(t)}{F(t)},
 \end{equation}
for details see Gupta et al. (1998), Di Crescenzo (2000).\\

It has been shown by many authors that under certain conditions  the measure defined in (3) characterizes some specific distributions, for details refer to Nanda and Paul (2006), Kundu et al. (2010). As mentioned earlier, Ebrahimi (1996) has shown that the measure of residual entropy (2) characterizes the distribution function uniquely. A result that in general, the measure of past entropy also characterizes the distribution function uniquely has not been studied so far. In this communication, we  study a result to this effect in the form of theorem given below\\

\textit{\textbf{Theorem}} \textit{Let $X$ and $Y$ be two random variables with absolutely continuous cdf $F$ and $G$, pdf $f$ and $g$, respectively having finite entropies.
 Assume that $\exists$ a constant $t_{0}$ such that $F(t_{0})=G(t_{0})$. Then the following two statements are equivalent\\
(i) $X$ is identical in distribution with $Y$.\\
(ii) $\bar{H}(X;t_{0})=\bar{H}(Y;t_{0})$.\\}
\textit{Proof:} (i) implies (ii) is obvious. We need to show only that (ii) implies (i). From (3) we have
\begin{eqnarray*}
  \bar{H}(X;t) &=& -\int_{0}^{t}\frac{f(x)}{F(t)}\log{\frac{f(x)}{F(t)}}dx .\end{eqnarray*}
It can be easily seen that
\begin{eqnarray*}
  \bar{H}(X;t)&=& 1-E\left[\log{{\phi}_{X}(X)}\mid{X<{t}}\right].\end{eqnarray*}
Hence
 \begin{equation}1-\bar{H}(X;t)=\int_{0}^{t}\frac{f(x)}{F(t)}\log{\frac{f(x)}{F(x)}}dx. \end{equation}
Using probability integral transformation $U=F(X)$, where $U$ is the standard uniform distribution, we get
\begin{equation}
\bar{H}(X;t)=\log{F(t)}-\frac{1}{F(t)}\int_{0}^{F(t)}\log{f({F}^{-1}(u))}du.
\end{equation}
We have \begin{equation}
\bar{H}(X;t_{0})=\bar{H}(Y;t_{0}).\end{equation}
Using (6) in (7) and using the fact $F(t_{0})=G(t_{0})$=$v$ (say), we get
\begin{equation}
\int_{0}^{v}\log{\left(\frac{f({F}^{-1}(u))}{g({G}^{-1}(u))}\right)du}=0.\end{equation}
So, $f({F}^{-1}(u))$=$g({G}^{-1}(u))$, $\forall{u}$ $\varepsilon{(0,1)}$. Using $\frac{d}{dt}(F^{-1}(t))=\frac{1}{f(F^{-1}(t))}$ and $F(t_{0})=G(t_{0})$ the result follows.\\\\
\textbf{References}\\
 1. Di Crescenzo, A.  Longobardi, M., (2002). Entropy-based measure of uncertainty in past lifetime distributions. Journal Appl. Prob. 39, 434-440.\\
 2. Di Crescenzo, A.,(2000). Some results on the proportional reverse hazards model. Statist. Prob. Lett.50, 313-321.\\
 3. Ebrahimi, N.,(1996). How to measure uncertainty in the residual lifetime distribution. Sankhya A58, 48-56.\\
 4. Gupta, R.C., Gupta P.L., Gupta, R.D.,(1998). Modeling failure time data by lehman alternatives. Communication in Statistics: Theory and Methods 27(4), 887-904.\\
 5. Kundu, C., Nanda, A.K., Maiti, S.S., (2010). Some distributional results through past entropy. Journal of Stst. Plan. and Infer. 140, 1280-1291.\\
 6. Nanda, A.K., Paul, P., (2006a). Some properties of past entropy and their applications. Metrika 64, 47-61.\\
 7. Shannon, C.E.,(1948). A mathematical theory of communication. Bell System Tech. J. 27, 379-423 623-656.
\end{document}